\documentstyle[aps,prd]{revtex} 
\draft

\begin{document}
\title{%
\hbox to\hsize{\normalsize\rm November 1999
\hfil Preprint MPI-Pth/99-55}
\vskip 36pt
             Neutrino Flight Times in Cosmology}
\author{L.~Stodolsky}
\address{Max-Planck-Institut f\"ur Physik 
(Werner-Heisenberg-Institut),
F\"ohringer Ring 6, 80805 M\"unchen, Germany}
\maketitle
\begin{abstract}
If  neutrinos have a small but non-zero mass,
time-of-flight effects for neutrino bursts from distant sources can
yield information  on the large-scale geometry of the universe,
the  effects being proportional to  $\int a~ dt$,
where $a$ is
the cosmological expansion parameter. In principle absolute
physical determinations of the Hubble constant and the acceleration
parameter are possible. Practical realization of
these possibilities would depend on neutrino masses being in a
favorable range  and the future development of very
large detectors. 
\end{abstract}
\vskip2.0pc


The existence of a small but non-zero mass for the neutrino,
combined with very long flight
times from astronomical sources,  offers the possibility for some
fundamental investigations, such
as tests of the assumptions of relativity~\cite {tests}. There is
increasing evidence~\cite{kam} for the existence of a small
neutrino mass, and here we would like to
note the further possibility of
using such effects to study the parameters of cosmology. We 
require the detection of neutrino
bursts from events at cosmological distances. This appears very 
difficult and may never be
feasible. On the other hand we would have, along with such classic
measures as the dependence on distance of the apparent
brightness 
of ``standard candles'' or the angular size of ``standard
measuring rods'', a 
totally different method for investigating the large scale geometry
of the universe. The method
might be called ``physical'' as opposed to ``geometric'' and would
not be subject to questions
concerning evolutionary effects in the way that the ``standard''
candles and rods are. 

As opposed to an exactly massless particle, which
always travels on the light cone,
a neutrino with a small mass has a velocity deviating from the
speed of light  in a way
which depends on the cosmological epoch. Hence the delay in arrival
time of the neutrino
compared with a photon emitted in the same event, or the relative
arrival times among neutrinos
of different masses emitted in the same event, contains information
on the cosmological epochs
through which the neutrino has passed.  

 This is  seen in the standard FRW metric~\cite{kolb}  $ds^2=dt^2
-a^2(t)(d{\bf x})^2$, where we define $a(t)$ to be the expansion
factor of the universe normalized to its present value: $a(t)=
R(t)/R(now)$, so that $a(now)=1$. We proceed by finding  an
equation for the
 coordinate velocity $ dx^i/dt$, where $x^i$ is along the 
particle's flight direction. First we express $ dx^i/dt$ in terms
of  
$P^i(t)$,  the  spatial part of the contravariant four-momentum 
$m~dx^{\mu}/ds$.  From the definition
of the metric we have
 $a(t) dx^i/dt =[a(t) P^i(t)]/\sqrt{m^2+[a(t) P^i(t)]^2}$.
 
 Expanding for the relativistic case $P\gg m$ we obtain $a(t)~
dx^i/dt \approx
1-{1\over 2}m^2/[a(t)P^i(t)]^2$
To find $P^i(t)$, we   now make use of the fact that  the covariant
or   ``canonical
momentum''
$P_{i}$ is constant (since nothing
depends on the $\bf
x$-coordinate). Furthermore,  since the different kinds of momenta
are
 related through the metric tensor, they all become equal at
$t(now)$, where 
$a=1$. Hence we can identify the  constant covariant momentum as
$P(now)$, the momentum at the detector. 
 Thus from  $P^i =g^{ij}P_j=1/(a^2)P_i$, we obtain $P^i
=1/(a^2)P(now)$ . Thus we
finally have
\begin{equation} \label{v}
{dx\over dt}\approx {1\over a(t)}-a(t)
{1\over 2}  {m^2\over P^2(now)}.
\end{equation}
Or introducing $\Delta x$ for the difference in the $x$ coordinate
of two different particles of mass
$m_2$ and $m_1$ emitted in the same event

\begin{equation} \label{deltax}
{d(\Delta x)\over dt}\approx
 a(t){1\over 2} \big[{m_1^2\over
P_1^2(now)}- {m_2^2\over P_2^2(now)}\big] 
\end{equation}

  At the present epoch with $a=1$, $\Delta x$ is just the spatial
separation of
the two particles.
Integrating, we have for this separation, or in view of $ v \approx
c=1$ for the time difference
in arrival at a detector

\begin{equation} \label{integral}
\Delta t\approx \Delta x\approx \int a(t)~dt {1\over 2}
\big[{m_1^2\over P_1^2(
now)}- {m_2^2\over P_2^2(now)}\big].
\end{equation}

 A somewhat different effect follows   if we consider a single mass
state, or equivalently if the mass differences in question are
undetectably small. Given a non-zero mass there is  still a
time-of-flight effect due to the variation  of velocity with
energy.
This  
was  the original observation of Zatsepin~\cite{zat} at the
beginning of
this subject, namely that in a burst there is a range of energies
in general and therefore a non-zero mass for the neutrino leads  
to a dispersion of velocities, and so there is
a spreading  as the burst travels.  For two momenta $P(now)$ and
$P'(now)$ we have
  
\begin{equation} \label{spread}
\Delta t \approx \int
 a(t)~dt {1\over 2} m^2
\big[({1\over P(now)})^2- ({1\over P'(now)})^2\big].
\end{equation} 
Thus 
  the spreading effect is governed  by the same 
 cosmological factor $\int
a~ dt$ as the time delay.  
The integral over $t$ is from the time of emission to the
present (``$t(now)$'') and contains
information on the cosmology.
With the expansion~\cite{kolb} of $a(t)$ for recent epochs 
$ a(t)= 1+ H
[t-t(now)]-{1\over 2} q  H^2 [t-t(now)]^2+...$,  and the redshift
parameter $z=1/a -1=-H[t-t(now)]+(1+q/2) H^2 [
t-t(now)]^2+...$, we can give the results of the integration in
terms of the directly observable $z$, in a low $z$ approximation to
order $z^2$, in terms  of  
the present Hubble constant $H$  and acceleration parameter $q$:

\begin{equation}\label{delayz}
 \Delta t \approx {z\over H}\big[1-{3+q\over 2} z +... \big]
{1\over 2}
\big[{m_1^2\over P_1^2(
now)}- {m_2^2\over P_2^2(now)}\big]
\end{equation}
while for the burst spreading we have

\begin{equation} \label{z}
\Delta t \approx {z\over H}\big[1-{3+q\over 2} z +... \big]{1\over
2} m^2
\big[({1\over P(now)})^2- ({1\over P'(now)})^2\big].
\end{equation}

The first thing to notice about these formulas is that they provide
a direct physical  determination of $H$, the Hubble constant.
Knowing the neutrino mass or masses, the neutrino energy  and the
redshift of the source,  measurement of a $\Delta t$ determines
$H$. It would certainly be interesting to compare a determination
of $H$ in this ``absolute'' way with the classical astronomical
results.

Observe that it is not necessary to know the distance to the
source,--- or
rather a ``distance'' is determined through the particle velocities
and their $\Delta x$ at the detector. Thus in addition to other
classical distance measures in cosmology, like the ``luminosity
distance''~\cite{kolb}, we may say that  there is another, the
``neutrino'' or  ``flight time'' distance, $d_{\nu}$, defined
through $\Delta x= \Delta v~ d_{\nu}$, where 
 $\Delta v$ is ${1\over 2}
\big[{m_1^2\over P_1^2(
now)}- {m_2^2\over P_2^2(now)}\big]$, or ${1\over
2} m^2
\big[({1\over P(now)})^2- ({1\over P'(now)})^2\big]$.

The cosmic acceleration $q$ appears in the quadratic correction.
Here again,
knowing the neutrino masses, their energy, redshift $z$ of
the source, and the Hubble constant, we could---in principle---find
the acceleration parameter from just one event.
Recent reports~\cite{q} of a non-zero value of $q$ from
a ``standard candle'' method  using distant supernovas has
attracted
considerable attention. It would  be of the greatest
interest if a totally different method could confirm these results.

 Although we have carried out the above arguments
for the spatially flat ($k=0$) FRW metric, it may be verified that
due to the existence of a similar constant ``canonical momentum''
in the other, 
 spatially curved ($k=\pm 1$) FRW metrics,
 the same arguments go through with the same results.
  
 The most straightforward application of these ideas would be
to situations  where  bursts of photons ($m=0$) or
different neutrino mass states, arriving distinctly separated in
time, are detected.
Assuming the neutrino masses have been well established by the time
such detections become feasible, two well-measured events  whose
redshifts can be determined and~Eq.(\ref{delayz})
 would determine the Hubble constant and the acceleration
parameter.  Unfortunately, the observation of events with  the
different 
mass components well separated seems unlikely. Given that
we
expect neutrino masses in the eV range or less, we can express 
our basic flight time factor as

\begin{equation}\label{delay}
{(m/{\rm eV})^2\over 2 (P/{\rm GeV})^2}\approx   50 \mu {\rm
sec/Mpc}
\end{equation}

It appears that even at a thousand Mpc, a substantial part of the
way across the visible universe, we may only expect some $\rm msec$
delays. 
There is of course the chance, not yet completely excluded, that
the heavier neutrino masses are in the range of some tens of {\rm
eV},
which could lead to delays of minutes, for the most distance
sources
at  GeV energies.
Although $\mu$sec timing or better is well within the capabilities
of photon and neutrino detectors, the duration of the  production
event and its fluctuations  are likely to involve much longer time
scales. For example gamma-ray bursts or supernova neutrino bursts 
last
on the order of several seconds. So, unless heavier neutrinos
and/or lower energies are accessible, this will
 make  the direct observation of the  relative delay of different
neutrino mass states or the delay relative to a photon signal
difficult to extract~\cite{raffelt}.  A detailed analysis
involving high statistics, with sources at various distances and
theoretical modelling of the production event would probably be
necessary. 

In view of these probable difficulties,
 it is important to note that the pulse spreading  and related
effects  represented by~Eq.(\ref{z}) offer potentially observable
time-of-flight effects involving a cosmological factor, even if
distinct masses cannot be resolved. However, unless the event is
intrinsically very short, the modelling of the  event and the
assumption that a certain type of event is the same at different
epochs become necessary. We should stress that~Eq.(\ref{z}) only
represents travel time effects on the width of the burst, these
will in general have to be
folded with the red shift of the intrinsic time scale of the event
itself. Note, however, that the latter $\propto z$ only and do not
involve a quadratic term.

 One may even carry this kind of  speculation a step further and
consider the cosmology of neutrino oscillations. That is, we can
consider neutrino mass differences so extremely small that coherent
oscillation phenomena are possible over cosmological
distances~\cite{reinartz}. Although not very likely of practical
realization, the question involves  some interesting conceptual
issues  and will be presented separately~\cite{cos}.

It is of course intriguing and of conceptual interest that there
is a different, ``non-geometric'' way of investigating cosmology.
As a
practical matter,
because of the small rate of neutrino interactions and the
reduction of the flux with distance, 
 the only hope that these speculations might become
reality  would seem to  lie with the further
development of the extremely big, Km
size detectors. 
 Since  time-of-flight effects decrease with increasing particle
 energy, for the time delay and burst spreading effects, lower
neutrino
energies,  and  instrumentation of
  detection systems for lower neutrino energies, are favored.
Unfortunately detection 
cross sections are smaller at lower energy, so even larger
detectors would  seem to be called for.   On the other hand, for 
the hypothetical oscillation effects alluded to above, higher
energies seem to be indicated~\cite{cos}; there
cross section are larger and may tend to compensate for
the general decrease of flux with energy.

 A fascinating aspect of studying  cosmology   by
these neutrino-based techniques, should it ever be possible, is
that the ``look back
time'' is potentially much deeper than with optical methods.  A
neutrino burst---highly redshifted--- can reach us essentially
undisturbed after neutrino
decoupling,  some minutes into the Big Bang.  If such bursts exist
(say from highly overdense regions collapsing to black holes) 
 the present  methods could allow us to determine the cosmology
back to
these early epochs. At such large values of $z$, of course, the
above small $z$ approximations
would   have to be replaced by expressions involving the full
cosmological model, as   would also be necessary for other high $z$
observations. If even more weakly interacting and non-zero mass
but highly relativistic particles than the usual neutrinos exist
(say right-handed neutrinos) the ``look back time'' goes deeper but
the detection difficulties get worse.

The only  presently known candidates at present for our procedures
are the
photon and the
neutrinos. Naturally, if  heavier but neutral and sufficiently
long-lived objects should exist and be emitted in cosmic events,
(say the WIMP of dark matter searches) there would be
 a substantial enhancement for  time-of-flight
effects. I am thankful to V. Zakharov for  this last remark and G.
Raffelt for helpful comments on the manuscript.




\end{document}